\documentstyle[aps,prl,epsfig,floats]{revtex}

\def\gsim{\lower0.5ex\hbox{$\:\buildrel >\over\sim\:$}}
\def\lsim{\lower0.5ex\hbox{$\:\buildrel <\over\sim\:$}}
\def \n{\noindent}

\begin{document}


\twocolumn[\hsize\textwidth\columnwidth\hsize\csname @twocolumnfalse\endcsname

\title{$b$-Parity: counting inclusive $b$-jets as an efficient probe of
new flavor physics}
\author{S. Bar-Shalom$^a$ and J. Wudka$^b$}
\address{$^a$ Physics Department, Technion-Institute of Technology, 
Haifa 32000, Israel.\\
$^b$ Department of Physics, University of California, Riverside CA 92521.}
\date{\today }
\maketitle

\begin{abstract}
We consider the inclusive reaction
$\ell^+ \ell^- \to nb +X$ ($n$ = number of $b$-jets) 
in lepton colliders for which 
we propose a useful approximately conserved quantum 
number $b_P=(-1)^n$ that we call $b$-Parity ($b_P$). 
We make the observation that the Standard Model (SM) is 
essentially $b_P$-even since SM $b_P$-violating signals 
are necessarily CKM suppressed. In contrast new flavor 
physics can produce $b_P=-1$ signals whose only significant 
SM background is due to $b$-jet misidentification.
Thus, we show that $b$-jet counting, 
that relies primarily on $b$-tagging, 
becomes a very simple and sensitive probe 
of new flavor physics (i.e., of $b_P$-violation).
\end{abstract}

\draft
\pacs{PACS numbers: 11.30.-j, 14.65.-q, 13.90.+i, 12.90.+b }


\vskip1pc]

%

The Standard Model (SM), despite its enormous success, 
is believed to be the low-energy limit of a more fundamental 
theory whose nature will be probed by the next generation 
of colliders. New physics effects have been
studied in a variety of processes using model independent 
approaches~\cite{leff}, as well as within specific models~\cite{spec.models}.
All such investigations aim at providing a 
clear unambiguous signal for non SM physics effects.
In this letter we propose one such signal obtained through 
simple $b$-jet counting. The approach is best suited to 
lepton colliders but may be extended to hadron colliders, 
e.g., it applies to the Fermilab Tevatron $p \bar p$ collider 
to the extent that the sea $b$-quark content of the protons can be 
ignored. 

Consider the inclusive multiple $b$-jet production in 
$\ell^+\ell^-$ collisions
\begin{eqnarray}
\ell^+ \ell^- \to n \cdot b +X \label{nbx}~,
\end{eqnarray}
where $n$ denotes the number of $b$ and $\bar b$-jets in the 
final state (FS) and $X$ stands for non $b$-jets, 
leptons and/or missing energy; it is understood that this represents the
state after top-quark decay.

For the reactions (\ref{nbx}) we introduce a 
useful approximate symmetry we call $b$-Parity ($b_P$), defined as
\begin{eqnarray}
b_P=(-1)^n ~.
\end{eqnarray}
In the limit where the quarks mixing CKM matrix $V$ \cite{ckm} satisfies 
$ V_{3j}=V_{j3}=0$ for $ j \not=3 $, all SM processes are 
$b_P$-even since in this case the third generation quarks 
do not mix with the others, and this leads  to the conservation of the
corresponding flavor number. 
Given the fast top decay and since
${\rm Br}( t \to b W) \simeq 1 $, the experimentally-observed
flavor number is in fact carried only by the $b$-quarks. Therefore,
the measured quantum number reduces to the net number of detected $b$-quarks;
we find it convenient to use instead the derived quantity $ b_P $.

The only SM processes that violate this conserved
number necessarily involve the charged current interactions
and are, therefore, suppressed by the corresponding small off-diagonal CKM elements
$|V_{cb}|^2$, $|V_{ub}|^2$, $|V_{ts}|^2$ or $|V_{td}|^2$. 
As a consequence the irreducible SM
background to $b_P=-1$ 
processes (induced by new flavor physics beyond the SM)   
is strongly suppressed: {\em the SM is essentially $b_P$-even}.

In the following we will look for experimental signatures of $ b_P$-odd
physics within multi-jet events. We will assume that a sample with a
definite number of jets has been selected (we will use 2 and 4 jet samples)
and determine (within each sample) the experimental sensitivity needed to detect
-- or rule out -- new flavor physics of this type up to a
certain scale. In
contrast with other observables the determination of
$b_P$ within a sample with a fixed number of jets
relies primarily on the $b$-tagging efficiency and purity of the
sample used and not on the particular
structure of a given FS, nor does it require 
the identification of any other particle but the $b$. 
Thus, the main obstacle in 
this use of $b_P$ is the reducible SM background, due to
jet mis-identification.
This results from having a 
$b$-tagging efficiency $ \epsilon_b$ below 1, and/or 
having non-zero probabilities $t_c$ and $ t_j $ of
misidentifying a $c$ or light jets for a $b$-jet, respectively.
This type of
background would of course disappear as $ \epsilon_b \to 1 $ and 
$t_{c,j} \to 0$, but even for the small value $t_c =0.1$ 
and high $b$-tagging  efficiency
$ \epsilon_b= 0.7$, can produce a significant number of 
(miss-identified) events in the detector.
Since for most experiments $ t_j $ is very small~\cite{opal.example},
the only relevant 
experimental parameters for this probe of new physics are $\epsilon_b$
and $t_c$.

Consider now the inclusive $b$ and $\bar b$-jet production process 
in (\ref{nbx}).\footnote{To be specific, we will
consider reactions in $e^+e^-$ colliders, but the method is clearly
extendible to muon colliders.} 
Focusing only on multi-jet FS, let $\sigma_{n m \ell}$ be the 
cross-section for   
\begin{equation}
e^+e^- \to n \cdot b + m \cdot c + \ell \cdot j \label{nml} ~,
\end{equation}
where $j$ is a light-quark or gluon jet and $c$ is a $c$-quark jet. 
Since our method does not require the detection of the
charge of the $b$, $n$ is the 
number of $b$ + $\bar b$-quarks and similarly $m$ and $\ell$ are 
the number of the corresponding jets in (\ref{nml}) irrespective 
of the parent quarks charges.

We denote by $t_c$ the $c$-jet mis-tagging probability 
(i.e., that of mistaking a $c$-jet for a $b$-jet) and by $t_j$ the light-jet 
mis-tagging probability (i.e., that of mistaking a light-jet or gluon-jet for 
a $b$-jet). Using these,   
the probability (or cross-section) for detecting precisely $k$ $b$-jets 
in the  reaction (\ref{nml}) is given by
\begin{eqnarray}
\bar\sigma_k = \sum_{u,v,w} && P_u^n P_v^m P_w^\ell 
\left[ \epsilon_b^u (1-\epsilon_b)^{n-u} \right]
\left[ t_c^v (1-t_c)^{m-v} \right] \nonumber \\
&&\left[ t_j^w (1-t_j)^{\ell - w} \right] \sigma_{ n m \ell} \delta_{u+v+w,k} 
\label{sigmabar} ~,
\end{eqnarray}
where $P^i_j = i!/j!/(i-j)!$.  

To experimentally detect $b_P$-odd signals 
generated by new physics one should simply 
measure the number of events with an odd number of 
$b$-jets in the FS. In particular, for the reaction (\ref{nml}),  
we define $N_{k,J}$ to be the number of events 
with $k$ (taken odd) $b$-jets in a FS with 
a total of $J$ jets.
The sensitivity of $N_{k,J}$ to $b_P$-violating 
new physics is determined  
by comparing the theoretical shift due to the underlying 
$b_P=-1$ interactions 
with the expected error ($\Delta$) 
in measuring the given quantity. 
Thus,
requiring a signal of at least $N_{SD}$ standard deviations, we have
\begin{eqnarray}
\left| N_{k,J} - N_{k,J}^{\rm (SM)} \right| \geq N_{SD} \Delta \label{nsd}~.
\end{eqnarray}
We will include three contributions to $ \Delta $ which we combine in
quadrature:
$\Delta^2 = \Delta_{\rm stat}^2 + \Delta_{\rm sys}^2 + \Delta_{\rm theor}^2$,
where $\Delta_{\rm stat}=\sqrt {N_{k,J}}$ is 
the statistical error, $\Delta_{\rm sys}=N_{k,J} \delta_s$ is a systematic 
error and $\Delta_{\rm theor}= N_{k,J}
\delta_t$ is the theoretical error in the numerical integration of the
corresponding cross sections. 
The quantities $ \delta_{s,t} $ denote the statistical and
theoretical errors per event; $ \delta_s $ is estimated using experimental
values from related processes (eg. $R_b$ measurements), $ \delta_t $ is
derived from the errors in the Monte Carlo integration used in
calculating the various cross sections.

There are various types of specific 
models beyond the SM (e.g., 
multi-Higgs models, supersymmetry, etc.) 
that can alter the SM prediction 
for the cross-section of reaction (\ref{nml}).
In this letter we will
take a model-independent approach in which
we investigate the limits that can be placed on the scale $\Lambda$ of
a new short-distance theory that can generate flavor
violation, and which we parameterize using an
effective Lagrangian~\cite{aewpaper} 
\begin{eqnarray}
{\cal L}_{eff} = \frac{1}{\Lambda^2} \sum_i f_{i} {\cal O}_{i}  +
O(1/\Lambda^3)
\label{leff}~,
\end{eqnarray}where ${\cal O}_{i}$ are mass-dimension 6 gauge-invariant
effective operators (some of which 
may have new flavor dynamics)~\footnote{We assume
there are no significant lepton-number violation effects at scale $ \Lambda $
that would generate dimension 5 operators.}, and
$f_i$ are coefficients that can be estimated using naturality arguments\cite{aewpaper}.

As a concrete example that clearly illustrates the significance of
$b_P$, we consider the effects of the
$b_P$-odd effective four-Fermi operator
\begin{eqnarray}
{\cal O} = \left( \bar \ell \gamma^\mu \ell \right) 
\left( \bar q_i \gamma_\mu q_j \right) \label{fourfermi}~,
\end{eqnarray}
where $\ell$ and $q$ are the SM left-handed lepton and 
quark $SU(2)_L$ doublets and $i,j=1,2~{\rm or}~3$ label the 
generation.  
This operator gives rise to contact $e^+e^- t \bar c$ and 
$e^+e^- b \bar s$ vertices  (and their charged conjugates). It
can be generated, for example, by an exchange
of a heavy boson in the underlying theory 
(see \cite{aewpaper}). 
Although our method applies to any $b_P=-1$ process,
 in what follows 
we will investigate the effects of
(\ref{fourfermi}) on the reaction (\ref{nml}) as an illustration. In particular,  
on $N_{1,2}$ (i.e., 1 $b$-jet signal in a 2-jet sample, $J=n+m+\ell=2$) and 
on $N_{1,4}$ and $N_{3,4}$ (i.e., 1 and 3 $b$-jet signals in a 4-jet sample, 
$J=n+m+\ell=4$). 

Consider first the $2$-jet sample case: in the limit 
$m_q=0$ for all $q \neq t$, the only relevant cross-sections 
are $\sigma_d = \sigma(e^+ e^- \to d \bar d)$, 
$\sigma_u = \sigma(e^+e^- \to u \bar u)$, where $d=d,~s,~b$ and 
$u=u,~c$, that are generated 
by the SM, and $\sigma_{bs} = \sigma(e^+e^- \to b \bar s) = 
\sigma(e^+e^- \to \bar b s)$ generated by the 
$eebs$ contact term.
These cross-sections are calculated by means of the CompHEP
package \cite{comphep}, in which we implemented the Feynman rules
for the $e^+e^-b \bar s$ and $e^+e^- t \bar c$ vertices generated 
by the operator (\ref{fourfermi}).
Using (\ref{sigmabar}), we get the following cross-section 
for the 2-jet events, one of which is identified as a $b$-jet 
($\bar\sigma_1$ with $J=n+m+\ell=2$)

\begin{eqnarray}
\bar\sigma_1 &=& P^2_1 \left[ \epsilon_b (1-\epsilon_b) + 
2 t_j (1-t_j) \right] \sigma_d \cr
&& + P^2_1 \left[ t_c (1-t_c) + 
t_j (1-t_j) \right] \sigma_u \cr
&& + 2 (P_1^1)^2 \left[ \epsilon_b (1-t_j) + 
t_j (1-\epsilon_b) \right] \sigma_{bs}
\label{sig1} ~
\end{eqnarray}
   
\n that is used to calculate $N_{1,2}$.
In Table \ref{tab1} we give the largest $\Lambda$
(the scale of the new $b_P=-1$ physics) that can be probed or excluded
at the level of
3 standard deviations ($N_{SD} = 3$),
derived using (\ref{nsd}), for the three representative $b$-tagging
efficiencies of $25\%,~40\%$ and $60\%$ and fixing the $c$-jet and light-jet 
purity factors to $10\%$ and $2\%$, respectively.\footnote{Note 
that the limits 
derived here and throughout the rest of the paper 
assume $|f|=1$ [see (\ref{leff})]. Alternatively, they can be 
interpreted as limits on  
$\Lambda/\sqrt {|f|}$.} 
Results are given for three collider scenarios:
$\sqrt s = 200$ GeV 
with $L=2.5$ fb$^{-1}$, $\sqrt s = 500$ GeV 
with $L=100$ fb$^{-1}$ and $\sqrt s = 1$ TeV 
with $L=200$ fb$^{-1}$.
Both the systematic 
error $\delta_{s}$ and the theoretical uncertainty $\delta_{t}$
are assumed to be $5\%$.  Also, an angular 
cut on the c.m. scattering angle of $|\cos\theta| < 0.9$ is imposed on each 
of the 2-jet cross-sections in (\ref{sig1}).
As expected, we see from Table \ref{tab1} that 
the sensitivity to the new flavor physics induced by the 
four-Fermi interaction increases with the $b$-tagging efficiency.

\begin{figure}[htb]
\psfull
\begin{center}
  \leavevmode
\epsfig{file=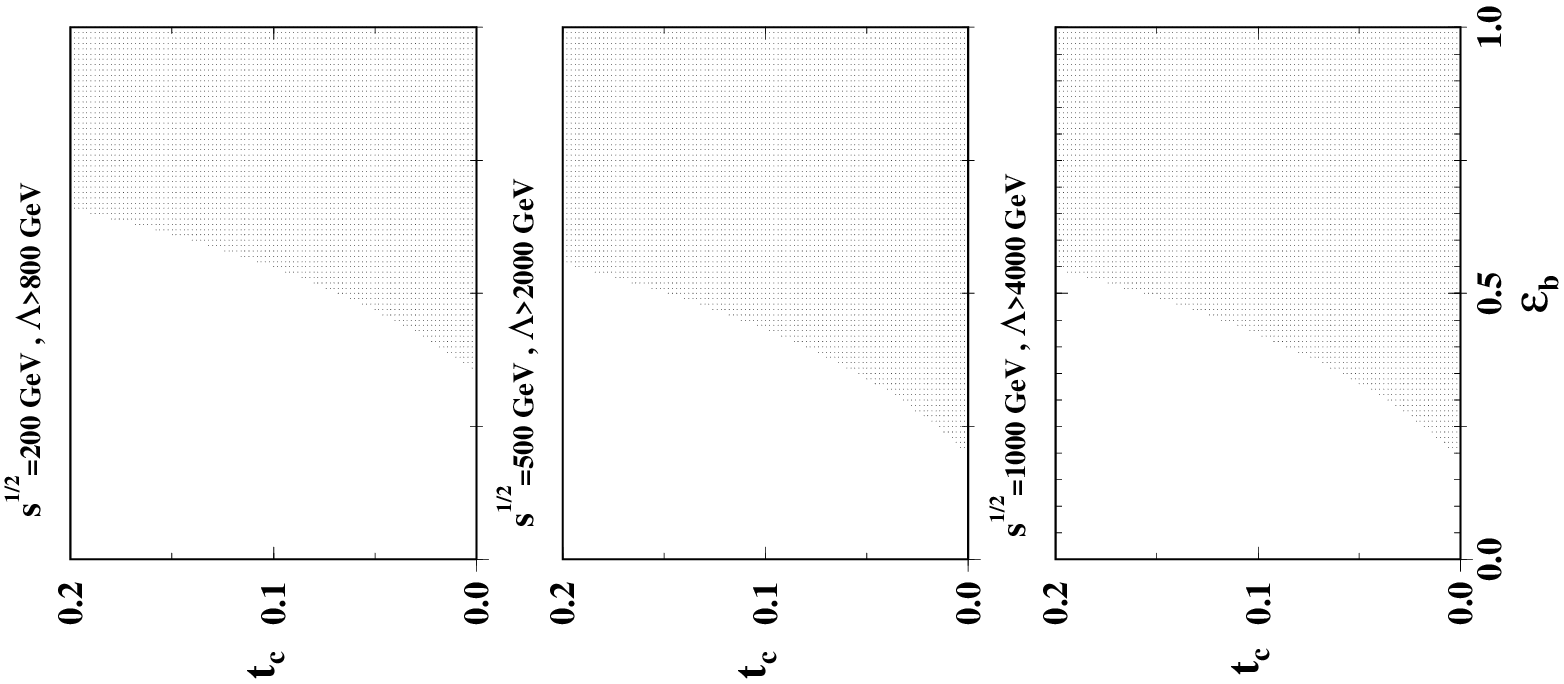,height=13cm,width=18cm,bbllx=0cm,bblly=2cm,bburx=22cm,bbury=25cm,angle=270}
 \end{center}
\caption{\emph{The shaded area denotes the region
of the $\epsilon_b - t_c$ plane  that can probe $\Lambda$
at the $3$ standard deviation level
(or higher). Results were obtained for three collider scenarios
using $N_{1,2} $ as an observable and assuming $\Lambda < 4 \sqrt s$.
We took $\delta_s=0.05$, $\delta_t=0.05$ and $t_j=0.02$ (see also text).
}}
\label{fig1}
\end{figure}

In Figs.~\ref{fig1} we show the regions
in the $\epsilon_b -t_c$ plane (enclosed in the dark areas) where the
flavor physics parameterized by (\ref{fourfermi}) can be
probed or excluded at the 
3 standard deviation level (or higher); as an illustration we chose
$\Lambda = 4 \sqrt s$ ($\sqrt{s}$ denotes the collider CM energy)
for the collider scenarios mentioned above. 
The calculation  was done using (\ref{nsd}) for $N_{1,2}$ 
with $\delta_s=\delta_t=0.05$, $|\cos\theta| < 0.9$, $t_j=0.02$.
Evidently, $\Lambda$ as large as four times the c.m. energy of any 
of the three  
colliders may be probed or excluded even for rather small 
$b$-tagging efficiencies; typically $\epsilon_b \gsim 25\%$ will 
suffice as long as the purity factors 
(in particular $t_c$  being the more problematic one)
are kept below the $10\%$ level.

\begin{table}
\caption{\emph{$3\sigma$ limits on $\Lambda$ (in TeV), the scale 
of the new $b_P=-1$ physics that generates the four-Fermi operator 
in (\ref{fourfermi}), as derived from $N_{1,2}$ in 2-jet events (see text).
}}
\label{tab1}

\bigskip

\begin{tabular}{|c|c||c|c|c|}
\hline
\multicolumn{5}{|c|}{$\delta_s=0.05$, $\delta_t=0.05$, $t_c=0.1$ and $t_j=0.02$}\\ \hline 
$\sqrt s$ & $L$        & $\epsilon_b=0.25$ & $\epsilon_b=0.4$& $\epsilon_b=0.6$\\
    (GeV) & (fb$^{-1}$)& $$                & $$              & $$              \\   \hline
200       & 2.5        & 0.68              & 0.74            & 0.81            \\   \hline
500       & 100        & 1.81              & 1.96            & 2.15            \\   \hline
1000      & 200        & 3.61              & 3.91            & 4.36            \\   \hline
\end{tabular}
\end{table}

For the 4-jet sample there are numerous processes that can 
contribute to $N_{1,4}$ and $N_{3,4}$. At the parton level the
4-jet events
may be categorized as follows: (1) events containing 2 quark-antiquark
pairs or one quark-antiquark pair and two gluons, 
$(q \bar q)( q^\prime {\bar q}^\prime)$, $(q \bar q) gg$, where 
both $q$ and $q^\prime$ denote any light quark 
($q,q^\prime \neq t$) including the case $q=q^\prime$. (2) events with 
two charged quark pairs: $(u \bar d) ({\bar u}^\prime d^\prime)$, 
where $u$, $u^\prime$ are either $u$ or $c$-quarks and 
$d$, $d^\prime$ are any of the down-type quarks,
excluding the states  $u=u^\prime$ and $d=d^\prime$
since these are induced in type (1) above.
(3) the 4 combinations $(b \bar s) gg$, $(b \bar b) (b \bar s)$, 
$(d \bar d) (b \bar s)$ 
and $(s \bar s) (s \bar b)$ 
(and the corresponding charged conjugate states) generated by 
the presence of the four-Fermi operator. It is worth noting that
the $eetc$ contact term also contributes through graphs containing a 
virtual top-quark exchange.

In order to get a reliable jet separation within the 4-jet sample, we use
the so-called Durham criterion \cite{durham}, that requires the quantities
$y_{ij}^D = 2 {\rm min}(E_i^2,E_j^2)(1-\cos\theta_{ij})/s$, 
where $E_i$ and $E_j$ are the energies of the particles $i$ and $j$
and $\theta_{ij}$ is their relative angle ($i \neq j = 1,\ldots,4$).
We evaluate all 4-jet cross-sections using the CompHEP 
package with the cuts $y_{ij}^D \geq y_{\rm cut}$
on all possible parton pairs $ij$ -- we 
present our numerical results for $y_{\rm cut}=0.01$.
In addition we neglect all quark masses except $m_{\rm top}$,
and the strong coupling constant $\alpha_s$ was evaluated to the
next-to-next-to leading order at a scale $Q$ equal to half the CM energy
for 5 or 6 active quark flavors depending on whether $Q<m_{\rm top}$ or
not respectively. For 6 flavors we used $ \Lambda_{QCD} =118.5$MeV (see
\cite{comphep} for details).

\begin{figure}[htb]
\psfull
 \begin{center}
  \leavevmode
  \epsfig{file=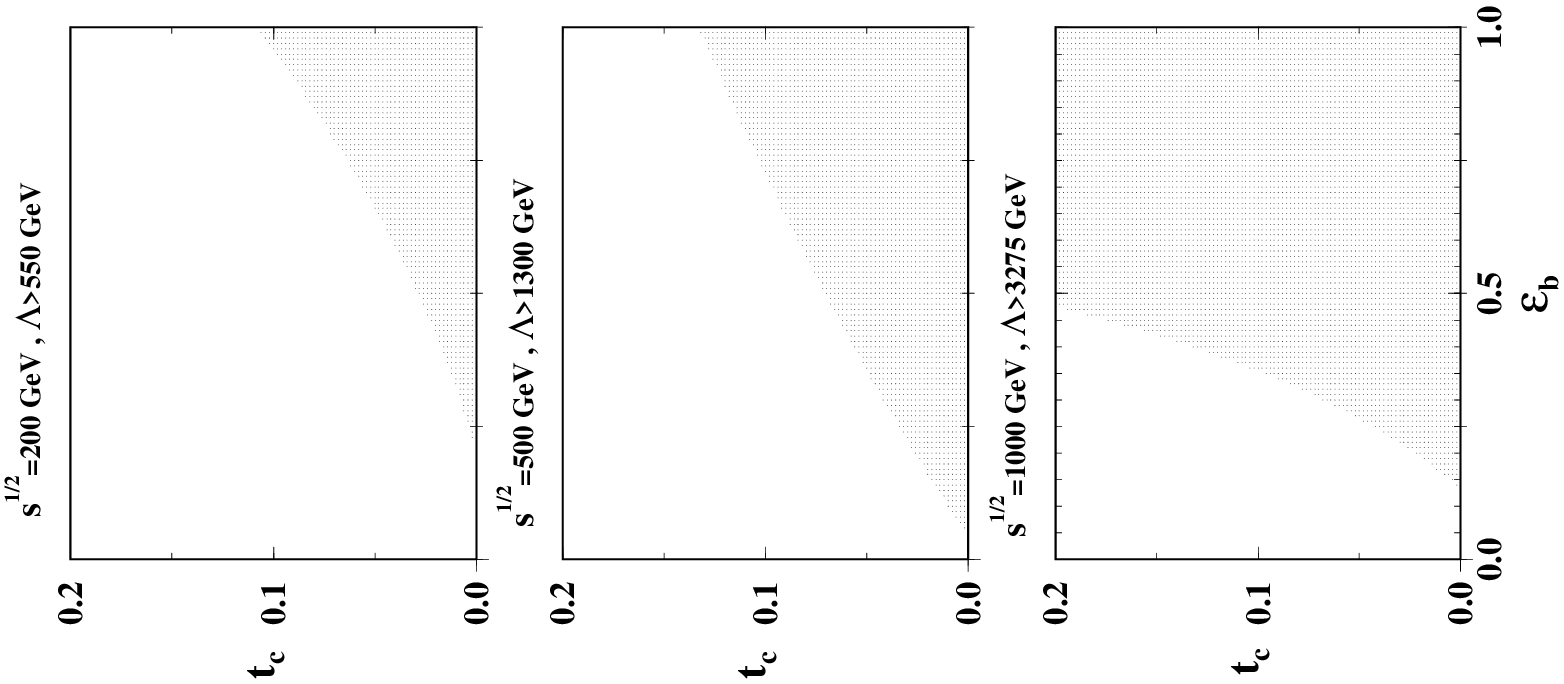,height=13cm,width=18cm,bbllx=0cm,bblly=2cm,bburx=22cm,bbury=25cm,angle=270}
 \end{center}
\caption{\emph{Same as in Fig. \ref{fig1} but for the 4-jet
case and using $N_{1,4}$. The values of $\Lambda$ being probed 
are indicated in the figure for any of the three collider scenarios. 
Here we take $\delta_t=0.1$ while $\delta_s$ and $t_j$ are kept 
as in Fig. \ref{fig1}.
}}
\label{fig2}
\end{figure}

The results for the 4-jet case, using $N_{1,4}$, 
are shown in Figs.~\ref{fig2} for a 200, 500 and 1000 
GeV colliders,
where, as in Fig.~\ref{fig1}, any value in the  
$\epsilon_b -t_c$ plane inclosed by the dark area will suffice 
for probing or ruling out (at $3\sigma$) the new four-Fermi operator in 
(\ref{fourfermi}) with a scale $\Lambda$ as 
indicated in the figure. As for the 2-jet sample, 
we take $t_j=0.02$ and a systematic error of $5\%$.
Using the results of the CompHEP Monte-Carlo integration
we estimate that our 
calculated 4-jet cross-sections 
are accurate up to the level of about 
$10\%$, accordingly we choose
$\delta_{t}=0.1$ in Fig. \ref{fig2}.

The 4-jet case is less sensitive to a $b_P$-odd signal 
induced by the four-Fermi operator in (\ref{fourfermi}). For example, 
we see that for a 500 GeV collider with $t_c \sim 0.1$ 
and $t_j=0.02$, 
a $b$-tagging efficiency of about $70\%$ will be needed in order 
to probe or exclude a value of $\Lambda \sim 1300$ GeV by measuring 
$N_{1,4}$, while a $40\%$ $b$-tagging efficiency will suffice 
for probing $\Lambda \sim 2000$ GeV using $N_{1,2}$ in the 2-jet 
sample. Equivalently, 
for a given value of $ t_{c,j}$,  higher $\epsilon_b$
will be required in the 4-jet measurement 
compared to the 2-jet one in order to detect a $b_P$-odd signal 
generated by the operator in (\ref{fourfermi}) for the same value of $ \Lambda $.
(this may be somewhat improved by reducing the theoretical uncertainties).
We also find that $N_{3,4}$ is less sensitive than $N_{1,4}$ 
to ${\cal O}_{\ell q}^{(1)}$ in (\ref{fourfermi}).

Though $N_{1,2}$ is more efficient for probing type of new flavor physics 
which generate the four-Fermi operator (\ref{fourfermi}) at low energies,
this is not necessarily a general feature: certain types of new
physics will not contribute to the 2-jet FS and must be probed using
the 4-jet sample. This is the case, for example, 
for an effective vertex generating a 
right-handed $Wbc$ coupling, that may alter the flavor structure 
of SM, and give rise to sizable $b_P=-1$ effects. Note, however,
that this refers only to the {\em exclusive} 2 and 4 jet samples, 
since such a 
$Wcb$ right-handed coupling 
will give rise to a $b_P$-odd signal in {\em inclusive} 2-jet reactions
such as $ e^+ e^- \to b +j + X $, where $j$ is a light jet.
The analysis of these events is, however, considerably more complex. 

We note that our cross sections include terms of order $ 1/\Lambda^4$
that will be modified by dimension 8 effective operators, which are in
general present in (\ref{leff}). Note, however, 
that such dimension 8 operators, if generated by the underlying 
high energy theory, are expected to give an
additional uncertainty of order $ (s/\Lambda^2)^2 $, below $3\%$ for
the results presented. In addition we note that the above analysis
assumes unbiased pure samples with a fixed jet number, the effects of
contamination from events with different jet number have not been
included.

Before we summarize we wish to note that the following issues 
need further investigation:

\begin{itemize}

\item Our $b$-jet counting method can be used to
constrain specific models containing 
$b_P=-1$ interactions. For example, 
supersymmetry with R-parity violation or 
with explicit flavor violation in the squark sector and/or  
multi-Higgs models without natural flavor conservation 
 can give rise to $t \to c,~ t \to u$ 
(or $b \to s,b \to d$) transitions, which 
may lead to sizable $b_P$-odd signals in leptonic colliders.

\item In leptonic colliders with 
c.m. energies $\gsim 1.5$ TeV, $t$-channel vector-boson fusion 
processes become important. At such 
energy scales, the SM $b_P=-1$ reducible background needs to be reevaluated. 
At the same time, the $V_1V_2$-fusion processes 
($V_{1,2} = \gamma,Z$ or $W$) give rise to a variety of new possible
$b_P=-1$ signals from new flavor physics (see e.g., \cite{mhdm}).

\end{itemize}

To summarize, we have shown that $b$-jet counting that relies
on $b$-tagging (with moderate efficiency in a relatively pure 
multi-jet sample) can be used to efficiently
probe physics beyond the SM. Reactions with $n$ final $b$-jets
can be
characterized through the use of the quantum number $b_P=(-1)^n$
that we called $b$-Parity. Due to small off-diagonal CKM matrix
elements, $b_P$ is conserved within the SM to very good accuracy;
it follows that the SM contributions to the above reactions are
$b_P$-even. Despite
the presence of a (reducible) background,  
due to reduced  $b$-tagging efficiency and sample purity,
we showed that our method is sensitive enough to provide
very useful limits on new flavor physics in a variety of scenarios (of which
two examples are provided) using realistic values for $ \epsilon_b $.

\acknowledgements
We would like to thank 
D. Atwood
R. Cole
W. Gary
R. Hawkings and
B. Shen
for illuminating comments and insights.
This research was supported in part by US DOE contract 
number DE-FG03-94ER40837(UCR).

\end{document}